\documentclass{elsart}

\usepackage{graphicx}

\usepackage{amssymb}
\begin{document}

\begin{frontmatter}



\title{ Nonlinear Optical studies of the Transient Coherence in 
the Quantum Hall  System}


\author[berk]{ K.M. Dani},
\author[crete]{ E. G. Kavousanaki}, 
\author[jt]{ J. Tignon}, 
\author[berk]{D. S. Chemla}, and  
\author[crete]{I. E. Perakis\corauthref{1}},
\corauth[1]{Corresponding author; Tel.: +30-2810-394259}
\ead{ilias@physics.uoc.gr}
\address[berk]{
Department of Physics, University of California, Berkeley, and 
Materials Science Division, Lawrence Berkeley National Laboratory, 
Berkeley, CA 94720, USA }
\address[crete]{
Department of Physics, University of Crete, and 
Institute of Electronic Structure \& Laser, Foundation
for Research \& Technology-Hellas, Heraklion, Crete, Greece}
\address[jt]{Laboratoire Pierre Aigrain,
Ecole Normale Sup\'erieure, F-75005 Paris, France}   

\begin{abstract}

We review recent investigations of the femtosecond 
non-linear optical response of the two--dimensional electron 
gas (2DEG) in a strong magnetic field. 
We probe the Quantum Hall (QH)  regime for filling 
factors  $\nu \sim 1$. Our focus is on the transient coherence 
induced via optical excitation and on its time evolution during 
early femtosecond timescales. We simultaneously study the 
interband and intraband coherence in this system by using a 
nonlinear  spectroscopic technique, transient three--pulse 
four wave mixing optical spectroscopy, and a many--body theory. We 
observe striking differences in the temporal and spectral profile 
of the nonlinear optical signal between a modulation doped 
quantum well system (with the 2DEG) and a similar undoped quantum 
well (without a 2DEG). 
We attribute these  qualitative differences to Coulomb 
correlations between the photoexcited electron--hole pairs and 
the 2DEG. We show, in particular, that intraband many--particle 
coherences assisted by the inter--Landau--level magnetoplasmon
excitations of the 2DEG dominate the femtosecond nonlinear optical 
responce. The most striking effect of these exciton--magnetoplasmon 
coherences is a large off-resonant four--wave--mixing signal in the case of 
very low photoexcited carrier densities,
not observed in the undoped system,  with strong temporal 
oscillations and unusually symmetric temporal profile. 
\end{abstract}
\end{frontmatter}


\section{Introduction}
\label{}

The quasiparticle concept is a cornerstone of modern condensed 
matter physics. The properties of many physical systems can be 
described, to first approximation, in terms of noninteracting 
quasiparticles and elementary excitations that may differ 
substantially from the strongly interacting bare electrons. In 
many cases, the quasi--static, thermodynamic, linear, and ground 
state properties do not depend critically on the residual 
interactions among the quasiparticles. On the other hand, these 
interactions determine the nonlinear 
optical dynamics \cite{chemla-1,chemla99}. 
For example, correlations among quasiparticles create quantum 
coherences in the system, 
which  lead to a new nonlinear optical signal, 
as well as  limit the lifetime of the 
collective excitations.
 Such dynamical effects govern the ultrafast 
nonlinear optical 
response measured in experiments such as transient 
wave-mixing and pump-probe 
\cite{chemla-1,schaf-book,haugbook,axt98,axt-04,rossi,perakis-rev}.
Coherent properties 
and their manipulation (i.e. coherent control) are central in many 
areas of physics and chemistry \cite{axt-04,bil,eit,lwi,inter-val}. 
They are currently the subject of intense research due to the many  
potential applications, e.g. in quantum coherent devices.   

In semiconductors, strong correlations and coherences can be mediated 
by the Coulomb and electron--phonon interactions. These effects are 
exacerbated in low dimensional nanostructures, where they can even 
lead to new quantum phases with novel transport and optical properties. 
An  example of such a low dimensional system is the cold two--dimensional 
electron gas (2DEG) subject to a strong magnetic field perpendicular 
to the confinement plane
 \cite{QHE1,QHE2,QHE3,QHE4}.
 This magnetic field creates discrete Landau 
levels (LL), which in the ground state are partially filled by electrons 
introduced by doping. 
The ratio of occupied states to LL degeneracy gives the filling factor
$\nu$. The LL degeneracy increases with magnetic field, and above a 
threshold value ($\nu \le 2$), the ground state electrons only occupy 
the lowest LL (LL0) states; all the higher LLs (LL1, $\cdots$) are 
then empty in the ground state. 
The coupling of the degenerate LL states by the 
Coulomb interaction results in a strongly correlated incompressible 
quantum liquid with collective charge and spin excitation modes. Examples 
of such collective excitations that play a central role here are the 
magnetoplasmon (MP) 2DEG excitations \cite{QHE1,QHE2,QHE3,QHE4,aron,eisen-01}. 
The strong exchange Coulomb interactions also stabilize a ground state 
with spin--$\uparrow$ polarized electron spins for certain integer values 
of $\nu$ or for certain fractional values $\nu =1/m$, where $m$ is an 
integer \cite{QHE2}. This paper discusses the crucial role 
in the transient optical properties of nonequilibrium 
Coulomb correlations 
between  photoexcited and 2DEG carriers 
during  very early femtosecond time scales. 

In addition to creating many--particle coherences and correlations,
the interactions among quasiparticles destroy coherence and phase
relations within short time intervals (dephasing or decoherence).
By probing the above Quantum Hall (QH) system with
very short optical pulses during time scales shorter than its dephasing 
time, one can observe coherent quantum mechanical effects not accessible 
with other experiments. In the very early temporal regime, the interactions 
among quasiparticles should be viewed 
as quantum mechanical interference phenomena, and 
well--established pictures, such as the semiclassical Boltzmann picture 
of dephasing and relaxation, must be revisited 
\cite{chemla-1,chemla99,schaf-book,haugbook,axt-04,rossi}. 
Indeed, the underlying assumption 
behind the above approximations is that the duration of the scattering 
and interaction processes is shorter than the time interval within which 
we probe the system. However, ultra-short pulses give access to timescales 
shorter than the characteristic interaction times, determined both by the 
time it takes for an exciton to dephase and by the time it takes the cold 
2DEG to react to the photoexcited excitons (X). The homogeneous dephasing 
times of the two lowest LL excitons in the QH system have been measured 
with ultrafast four--wave--mixing (FWM) spectroscopy. They range from a 
few picoseconds (LL0) to a few hundred femtoseconds (LL1). The reaction 
time of the 2DEG  is comparable to the period of its low 
energy collective excitations. The period of the lowest inter--LL MP
collective modes \cite{QHE1,QHE2} is $T_{MP} = 2 \pi \hbar/\Omega_{M}$, 
where $\Omega_{M} \sim 15-20$ meV is the MP excitation energy, which is of 
the order  of a few hundreds of femtoseconds. Thus, $\sim$100~fs optical 
pulses provide access to the quantum kinetic regime of the QH system.

Among the different nonlinear optical techniques, femtosecond 
FWM spectroscopy is well suited for studying coherent dynamics 
\cite{chemla-1,Shah99}. It has been used to demonstrate that, in undoped 
semiconductor quantum wells (QW), this dynamics 
is dominated by Coulomb interactions
\cite{chemla-1,chemla99,schaf-book,haugbook,axt98,axt-04,rossi}. 
Fluctuations beyond the Random Phase 
Approximation (RPA) generate a two--pulse FWM signal with a distinct time 
dependence \cite{chemla-1,chemla99}. In particular, 
the Pauli blocking (Phase Space 
Filling, PSF) effects \cite{Shah99} do not contribute during negative time 
delays, where exciton--exciton (X--X) interactions dominate in undoped QWs 
\cite{chemla-1,chemla99}. The time--dependent Hartree--Fock treatment 
of X--X interactions \cite{schaf-book} predicts an {\em asymmetric} 
FWM temporal profile, with a negative time delay signal decaying twice as 
fast as the positive time delay signal \cite{chemla-1,chemla99,Shah99}. 
The observation of strong deviations from this asymmetric  temporal 
profile was interpreted as the signature of X--X correlations in undoped 
QWs \cite{chemla-1,chemla99}.

Similar to undoped QWs, the interband  absorption spectrum of a 
modulation doped QW subject to a perpendicular magnetic field is dominated 
by discrete LL peaks \cite{from-02-prl,from-02-prb}. Despite 
some similarities however,
there are large qualitative differences in the spectral 
and temporal profile of the nonlinear optical signal stemming from the 
different  ways in which doped and undoped QWs respond to ultrafast 
excitation. In undoped QWs, the lowest electronic excitations are high 
energy interband transitions that react almost instantaneously to the 
presence of photoexcited carriers \cite{louie}. The ground state can then 
be considered as rigid, providing only the band structure and dielectric 
screening. Consequently, Coulomb correlations only occur among 
{\em photoexcited} carriers. The role of such interactions can 
be  analyzed by using theories such as the dynamics--controlled truncation 
scheme (DCTS) \cite{axt98,axt-04,Axt}, the correlation expansion \cite{rossi}, 
or the Keldysh Green function technique \cite{schaf-book,haugbook}.  
In contrast, in doped QWs, the presence of a 2DEG leads to strong Coulomb 
correlations in the ground state itself, which result in long-range charge 
and spin order at sufficiently low temperatures and to collective electronic 
excitations. As a result, the DCTS assumptions break down \cite{Axt}. The 
2DEG can respond unadiabatically to the interactions with the photoexcited 
Xs by emitting low energy  electronic excitations. 
One must distinguish such shake--up processes 
from those involving  the nonequilibrium photoexcited 
electron gas. The differences in the physics of the doped and undoped 
systems manifest themselves strongly in the ultrafast nonlinear optical response.

The study of the ultrafast nonlinear optical dynamics  of the QH system 
transcends across the boundaries  
of two  communities  largely disconnected up to 
now. Indeed, the transient optical properties of this system are governed 
by (i) the interband (X) excitations (with the 2DEG at rest), which consist 
of 1~$e$--$h$, 2~$e$--$h$, $\cdots$ pairs created in the LLs (studied by the 
nonlinear optics community), and (ii) the intraband 2DEG excitations (with 
unexcited QW and full valence band), e.g. the 1--MP, 2--MP, $\cdots$ and 
incoherent pair excitations (studied by the QH community). The ensemble of 
states that determine the nonlinear optical spectra 
to $(2\ell-1)$--th order in the optical field 
consist of products of 
up to $\ell$ $e$--$h$ pairs and $n$ 2DEG excitations. 
One can then 
draw an analogy with the X+phonon states 
that determine the ultrafast optical dynamics
in undoped semiconductors \cite{haugbook,axt98,rossi,Axt}. However, there 
are some important differences. In the QH system, the
2DEG excitations 
are electronic in nature, and therefore subject to Pauli correlations with 
the photoexcited Xs, while the ground state electrons are strongly 
correlated. On the other hand, in the 
undoped system, the X  operators commute with the collective 
excitation (phonon) operators, while the ground state correlations can be 
neglected. Thus 
the theoretical formulations used to study the nonlinear optical 
response in undoped semiconductors must be extended in order to 
treat  correlations in the doped system. 

Recent time--resolved FWM experiments  probed for the first time 
the coherent regime 
of the 2DEG in a magnetic field and opened a new field of non-equilibrium 
Quantum Hall physics 
\cite{from-02-prl,from-02-prb,from-99,kara,per-03,per-dan-ssc,per-dan-pss,schu,from-02-ph,dani,chem-phys}. The purpose of 
this article is to review the recent 
developments in this emerging field. We first briefly discuss the linear 
absorption and two--pulse FWM experiments that probe the exciton dephasing. 
There are striking differences when compared to an undoped sample. The 
inter--LL MP excitations, which provide a dynamical coupling between the 
LL0 and LL1 excitons, are responsible 
for these differences. We then focus on recent 
three--pulse FWM experiments, which give access to two different time 
delays. The FWM signal along one of these time delays probes the coherent 
inter--band dynamics of the QH system and reproduces the two--pulse FWM 
results. Measuring along the other time delay axis gives access to the 
intra--band dynamics of the QH system, 
which in the early coherent 
regime is governed 
by X$\leftrightarrow$X+MP, 
X$\leftrightarrow$X, and MP transient 
coherences. For short time delays along this 
new axis, the experiment reveals strong oscillations with a period given 
by the inverse LL0--LL1 energy splitting. For longer time delays, there is 
a slow rise of the three--pulse FWM signal on a timescale of several 
picoseconds, which only occurs for very low photoexcitation and is absent 
in the undoped QW. We also review a microscopic many--body theory that was 
developed in order to study the ultrafast coherent dynamics 
of strongly correlated systems
and interpret 
the experiments. By comparing to this theory we identify the physical 
origin of the observed oscillations. 
The combined experimental and 
theoretical investigations 
reviewed here  point out the important 
role of X$\leftrightarrow$X+MP coherences and correlations, 
created by the nonlinear photoexcitation and the 
long range order of the QH system. 

\section{Experimental results} 

The experimental results discussed here were obtained by performing 
linear and non-linear optical measurements on a modulation doped QW, 
whose active region consisted of 10 periods of a 12 nm GaAs well and 
a 42 nm Al$_{0.3}$Ga$_{0.7}$As barrier, the central 12 nm doped with Si. 
The sample was antireflection coated and mounted on sapphire windows 
for transmission measurements. The doped carrier density was 
$2.1\times 10^{11}$ cm$^{-2}$ and the low temperature mobility 
$\sim 10^5$ cm$^2$/Vs. The sample was kept in a magneto-optic cryostat 
at a temperature of $1.5-4^\circ$K. A perpendicular magnetic field 
($B=0-12$ T) was applied along the growth direction of the QW. The 
measurements in the above doped system were compared to those in an 
undoped sample (without a 2DEG) with similar well and barrier sizes.
The qualitative differences in the 
spectral and temporal profile of the linear and non--linear optical 
spectra of the two systems must thus be attributed to the 2DEG.
In order to make the interpetation easier, 
$\sigma_{+}$ circularly polarized pulses where used. In this case, 
only one transition is allowed by the selection rules, which photoexcites
spin--$\downarrow$ electrons. 
In this section we discuss 
the experimental linear absorption and transient  two-- and three--pulse FWM 
spectra in the two systems. 

\begin{figure}[t]
\begin{center}
\includegraphics[width=0.8\textwidth]{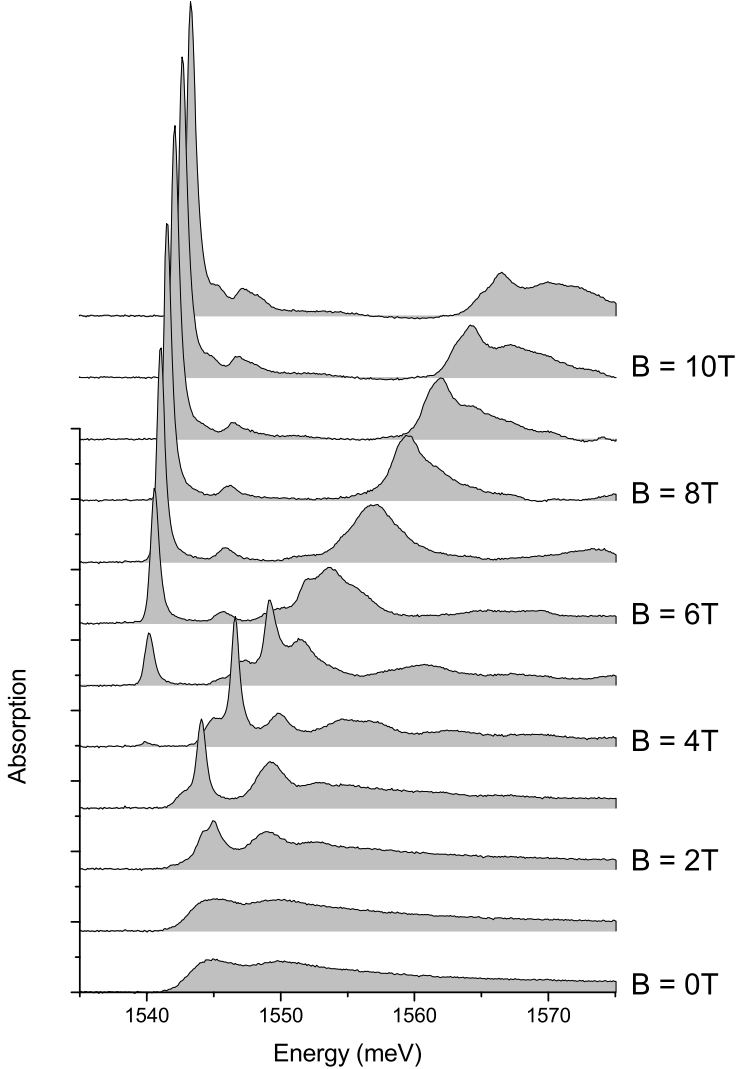}
\end{center}
\caption[]{
$\sigma_+$ linear absorption of the doped QW for different magnetic 
fields. One sees absorption into the two lowest LLs. The onset of 
absorption into LL0 (at 4.3 T, $\nu =2$) is accompanied by a sudden 
increase in the linewidth of the LL1 peak.
}
\label{LinAbsDoped}
\end{figure}

\subsection{Linear Absorption}

\begin{figure}[t]
\begin{center}
\includegraphics[width=0.8\textwidth]{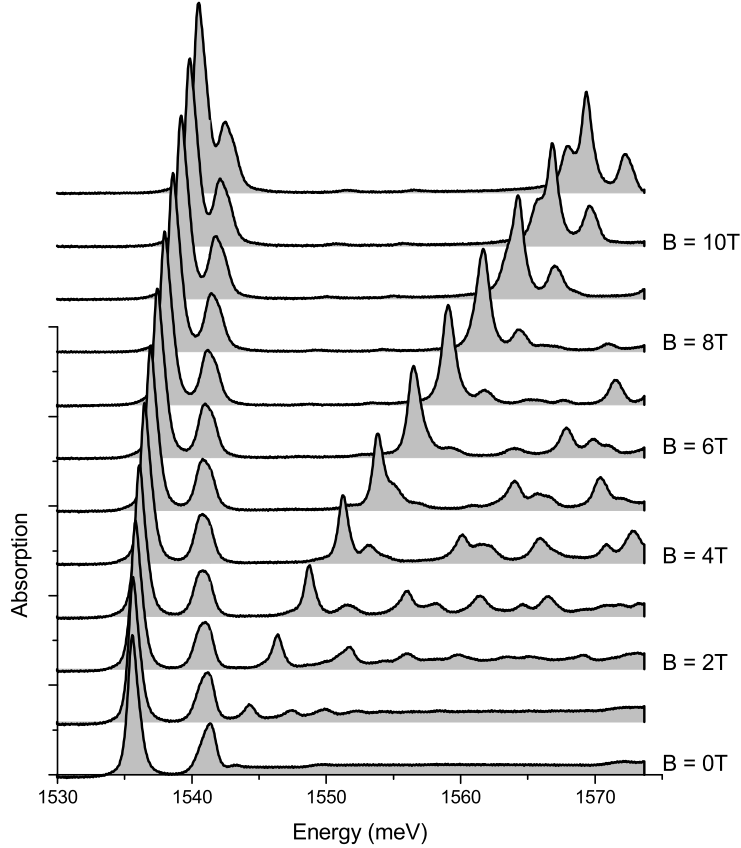}
\end{center}
\caption[]{
$\sigma_+$ linear absorption for the undoped QW. There is no 
significant change in linewidth of the LL0 or LL1 peaks versus 
magnetic field.
}
\label{LinAbsUndoped}
\end{figure}

We already see tell-tale signs of the important 
role of the cold 2DEG in the linear absorption spectrum \cite{from-02-prb}. 
Figs. \ref{LinAbsDoped} and 
\ref{LinAbsUndoped} compare the $\sigma_+$ linear absorption spectra 
of the modulation doped QW with those of the undoped QW for various 
magnetic fields. For both samples, one clearly sees the two  peaks 
that are due to absorption into the two lowest LLs (LL0 and LL1). In the 
doped sample, LL0 absorption is only possible when the filling factor 
$\nu$ is smaller than two, in which case empty LL0 states are available 
($B \ge 4.3$ T in Fig. \ref{LinAbsDoped}). On the other hand, in the 
undoped QW, one observes a sharp LL0 peak down to small magnetic fields. 
The striking feature in the doped sample is the coincidence between a large
increase in the linewidth of the LL1 peak and the onset of 
absorption into LL0. One sees no such increase in the LL1 linewidth 
in the undoped QW (Fig. 2). At the same time, the LL0 linewidth is not 
very sensitive to the filling factor $\nu$. We conclude that the increased 
LL1 linewidth results from an increase in the dephasing rate of the LL1 
magnetoexciton once empty LL0 states become available for scattering. 
For $\nu <2$, one expects scattering of the LL1 magnetoexciton (X$_1$) 
into a LL0 magnetoexciton (X$_0$) accompanied by the emission (shake--up) 
of inter--LL
MP excitations of the 2DEG. 
 Noting that the 
LL0$\rightarrow$LL1 magnetoplasmon is close in energy to the LL0--LL1 
magnetoexciton splitting, we expect that this X$_1 \rightarrow$X$_0$ + 
MP scattering process can be very efficient in limiting the 
lifetime of the LL1 exciton. In contrast, the LL0 exciton lifetime
is not affected much by the inter--LL MPs since there are no states 
available for resonant scattering. 
 The theoretical calculations 
discussed below confirm this scenario. 
The dynamics of the X$_1\rightarrow$X$_0$ + MP scattering process 
is studied here in full detail
by looking at the spectral and temporal profiles of the two-- and 
three--pulse FWM signals and by comparing between the doped 
and undoped QWs.
The 
inter--LL MPs play a crucial role in the transient nonlinear 
optical response by providing a dynamical exciton 
coupling and by introducing different  dynamics 
of the LL0 and LL1
excitons.

\begin{figure}[t]
\begin{center}
\includegraphics[width=1.0\textwidth]{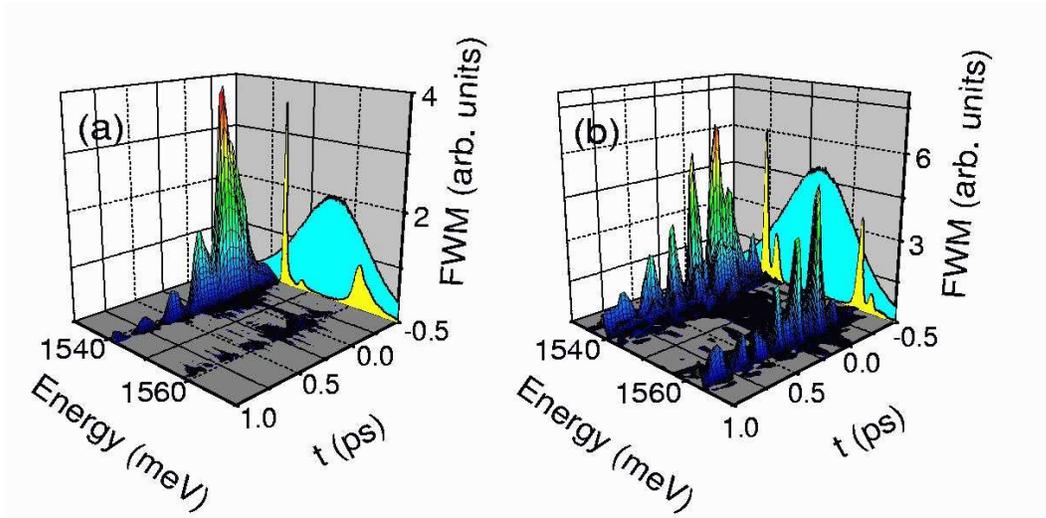}
\end{center}
\caption[]{
Comparison of 2-pulse FWM transient spectra when the two LLs are 
equally excited for (a) doped QW and (b) undoped QW.
The back panel shows the linear absorption and the pulse overlap 
with the two LL peaks.
} 
\label{EqualExcitationTwoPulse}
\end{figure}

\subsection{Two--pulse FWM}

In this section we discuss the two--pulse FWM experiments
\cite{from-02-prl,from-02-prb}, where the QH sample is excited  
with two pulses propagating along directions ${\bf k}_1$ and 
${\bf k}_3$ (named so for easier comparison with the three-pulse 
FWM experiments discussed in the next section). The 
FWM signal is measured in the direction 
$2 {\bf k}_1-{\bf k}_3$. We introduce a time delay $\Delta t$ 
(or $\Delta t_{13}$ for comparison with three--pulse FWM experiments) 
between the two pulses, where for negative delay pulse ${\bf k}_1$ 
comes first. Typically, the measurements were performed 
with weak photoexcitation intensity, keeping the total number of 
excited carriers under $2\times10^{10}$ cm$^{-2}$ or a tenth of 
the density of the 2DEG carriers. 
Only in this low excitation regime 
are the differences 
between the 
doped and undoped samples significant \cite{from-02-prb,dani}. 
 The FWM signal was spectrally resolved 
and the intensity measured as a function of wavelength (i.e. photon 
energy) as well as pulse time delay $\Delta t_{13}$. The spectral 
resolution allows one to separate out the contributions from the 
various LLs, which otherwise all contribute to the standard 
time--integrated FWM signal \cite{Shah99}. 
This spectral resolution and LL separation is important for 
identifying the physical mechanisms at work. 
Measurements under the 
same conditions were  performed on the undoped samples for comparison. 
The important criterion for this comparison to be meaningful, 
given the differences in the linear absorption between the two 
systems, was to excite the same number of electron-hole pairs into 
each LL with a given laser pulse.

We start by discussing the case of photoexcitation of almost equal 
numbers of LL0 and LL1 carriers. Several unusual features are immediately 
apparent in the FWM signal from the doped QW, as compared to the FWM 
signal from the similar undoped QW (see Fig. \ref{EqualExcitationTwoPulse}). 
The most striking is that, despite the equal excitation of both LLs, the 
doped QW shows a LL0 signal 35 times larger than the LL1 signal. In 
contrast, the undoped QW shows almost equal emission from both LLs, in 
proportion to the photoexcitation, as expected for a three--level system 
or RPA theory \cite{Shah99}. Importantly, although we see emission almost 
entirely from LL0, the doped QW FWM signal shows pronounced beats as a 
function of time delay $\Delta t_{13}$, with a period given by the 
inverse of the energy difference between the LL0 and LL1 absorption 
resonances. Strong beating for a single FWM resonance is a clear signal 
of non--Markovian dynamics. These beatings disappear 
in the case of LL1 photoexcitation 
discussed below. In contrast, the undoped QW shows beats 
from two emission peaks of similar strength, as expected  for 
a three--level system or from RPA theory \cite{Shah99}.

\begin{figure}[t] 
\begin{center}
\includegraphics[width=0.8\textwidth]{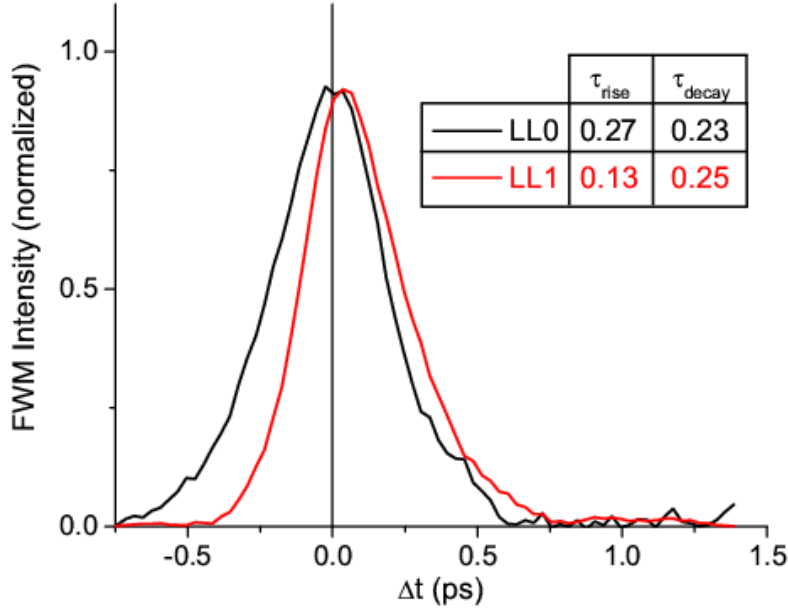}
\end{center}
\caption[]{
Normalized 2-pulse FWM emission from LL0 and LL1 when the LL1 is 
preferentially excited over LL0 (60:1 ratio).
} 
\label{LL1ExcitationTwoPulse}
\end{figure}

The picture is just as unusual when we tune the laser frequency to 
excite 60 times more carriers into LL1 than into LL0.
Despite the extremely weak LL0 photoexcitation, 
the LL0 FWM
signal in the doped QW is greatly enhanced: it is comparable to the 
LL1 signal despite the 
small 
PSF contribution at LL0. 
In contrast, in the undoped QW, 
there is almost no LL0 signal, as expected from the photoexcitation 
(60:1 for LL1:LL0). 
The large 
off--resonant LL0 
signal in the doped system   
can   come from LL1-LL0 coupling due 
to (i)  X-X interactions, (ii) inter-LL X
coherences,
and (iii) inter-LL coherences
assisted by the 2DEG excitations. 
 The first 
two also contribute in the undoped QW, where 
however  the LL0 signal is  small. 
Therefore, these contributions 
are weak, and we conclude that 
the LL0-LL1 coupling mainly comes from  
an inter--LL coherence assisted by the 
2DEG excitations. 

In addition to the transfer of FWM strength to 
LL0, the doped QW signal shows a very unique dependence on 
$\Delta t_{13}$ (Fig. \ref{LL1ExcitationTwoPulse}). 
According to the RPA theory \cite{weg-90}, the rise 
time of the $\Delta t_{13}<0$ FWM signal should be 1/2 the decay time 
for $\Delta t_{13}>0$ FWM, as measured in the undoped QW sample. This 
is also the measured result for the LL1 signal in the doped QW.
Surprisingly however, the LL0 signal is almost symmetric as a function of 
$\Delta t_{13}$, with comparable signals for $\Delta t_{13}<0$ and 
$\Delta t_{13}>0$ (see Fig. \ref{LL1ExcitationTwoPulse}). Such a large 
LL0 signal for $\Delta t_{13}<0$ can only be a result of correlation 
effects beyond the RPA \cite{chemla-1}. This doped QW effect is only seen 
for low photoexcitation intensity, which implies that the correlations 
are induced by the cold 2DEG.

\subsection{Three--pulse FWM}

In three pulse FWM, the sample was excited  with three 
100~fs, $\sigma_+$ pulses propagating along distinct directions ${\bf k}_1$, 
${\bf k}_2$, and ${\bf k}_3$. Pulses ${\bf k}_1$ and ${\bf k}_2$ (${\bf k}_3$) 
are separated by a time delay $\Delta t_{12}$ ($\Delta t_{13}$), where pulse 
${\bf k}_1$ arrives first for negative values of the delay. The FWM signal is 
obtained in the background--free direction ${\bf k}_1+{\bf k}_2-{\bf k}_3$. 
Using an interference filter, we spectrally resolve the signal so as to 
separate out the contribution from each LL. We then measure the intensity
from each LL as a function of the two above time delays. In particular, 
we measure along the $\Delta t_{12}$ axis ($\Delta t_{13}=0$) or the 
$\Delta t_{13}$ axis ($\Delta t_{12}=0$). Along the $\Delta t_{13}$ axis, 
three--pulse FWM results describe the interband polarization dephasing similar 
to two--pulse FWM in the direction $2{\bf k}_1-{\bf k}_3$. On the other hand, 
the three--pulse FWM signal 
along the $\Delta t_{12}$ axis reflects the dynamics 
of intraband coherences, e.g. coherences between X and X or X and X+MP states, 
or between the ground state and the MP states. By comparing the FWM temporal 
profiles as function of $\Delta t_{13}$ and $\Delta t_{12}$ during 
subpicosecond time scales we  arrive at a comprehensive picture of the 
dynamics of the interband and intraband coherences in the QH system. 

\begin{figure}[t]
\begin{center}
\includegraphics[width=1.0\textwidth]{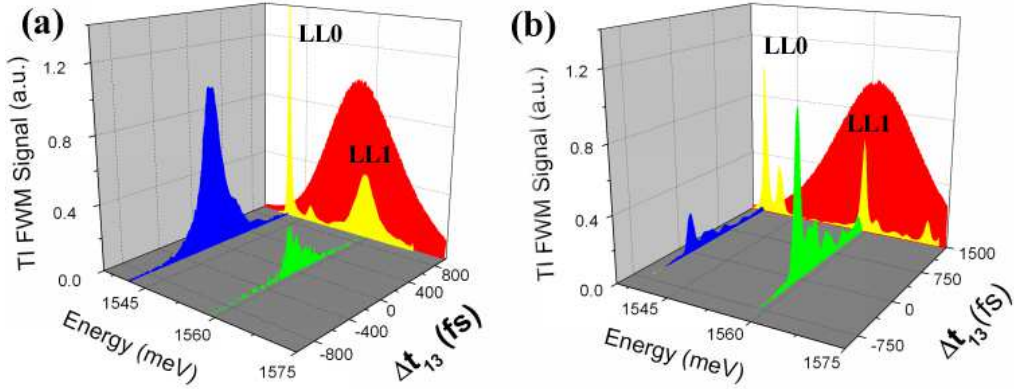}
\end{center}
\caption[]{
Comparison of the (a) doped QW and (b) undoped QW FWM spectra along 
the $\Delta t_{13}$ axis for LL1 photoexcitation. A temporally symmetric 
LL0 FWM signal dominates in the doped QW, while the LL1 
FWM dominates in the undoped QW.
The back panel shows the linear absorption and  pulse overlap 
with the two  peaks.
}
\label{LL1ExcitationDt13}
\end{figure}

The bulk of the three--pulse FWM measurements were performed at $B=7$ T, 
which corresponds to $\nu=1.3$ in this sample. Close to $\nu=1$, the results 
did not depend strongly on the filling factor.
The photo--excited carrier density (5$\times$10$^9$ cm$^{-2}$) was typically 
much lower than the density of the doped 2DEG carriers. Here we 
only focus on the
experiments for large excitation ratios of LL1 to LL0 (LL1:LL0 excitation 
ratio at least 10:1 in 
Fig. \ref{LL1ExcitationDt13}). The backpanel of Fig. \ref{LL1ExcitationDt13}
shows the overlap of the optical pulse with the LL0 and LL1 
peaks in linear absorption. 
In this case of predominantly LL1 photoexcitation, 
the LL0 signal can only arise from correlation effects.
Fig. \ref{LL1ExcitationDt13}a
 shows a large transfer of FWM 
signal strength 
from LL1 to LL0 in the doped QW, while 
Fig. \ref{LL1ExcitationDt13}b shows the FWM spectra for an undoped sample 
under similar excitation conditions for comparison. One does not see a 
large FWM  signal from LL0 in the undoped QW, as expected for LL1 
photoexcitation, while in the doped QW the LL0 signal clearly dominates.
 We also see in 
Fig. \ref{LL1ExcitationDt13}a 
the unusually symmetric temporal profile of 
the LL0 signal, similar to two-pulse FWM.

We now turn our attention to the large off-resonant LL0 signal 
along the $\Delta t_{12}$ axis. It is important to note that 
different time delays probe different physics in 
three--pulse FWM experiments. 
The measurements  along the $\Delta t_{12}$ axis provide 
new information about the dynamics of the intraband coherence 
that is not accessible with two--pulse FWM, 
while the $\Delta t_{13}$ axis probes  
the interband coherence. 
 For negative $\Delta t_{12}$, 
pulses ${\bf k}_1$ and ${\bf k}_3$ arrive together creating an 
intraband coherence, e.g. a X$\leftrightarrow$X+MP,
X$\leftrightarrow $X, or MP  coherence. 
For negative time delays, 
this coherence evolves for a time 
$|\Delta t_{12}|$, at which point it is probed by the arrival of 
pulse ${\bf k}_2$. Thus, during early femtosecond time scales,
the FWM signal versus $\Delta t_{12}$ 
reflects the dynamics of the intraband coherence.

\begin{figure}[t]
\begin{center}
\includegraphics[width=0.6\textwidth]{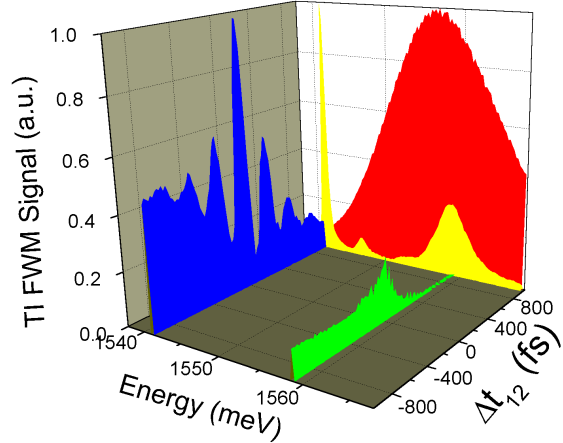}
\end{center}
\caption[]{
FWM spectra along the $\Delta t_{12}$ axis from the doped QW 
for large LL1/LL0 excitation. In the initial coherent regime, the 
LL0 signal displays strong oscillations.  
}
\label{LL1ExcitationDt12short}
\end{figure}

Fig. \ref{LL1ExcitationDt12short} 
shows the spectrally resolved FWM signal in the 
doped QW along the $\Delta t_{12}$ axis for low photoexcitation 
and low temperature. We observe strong oscillations in the off-resonant 
signal at LL0 as function of $\Delta t_{12}$. There are only minor
oscillations at LL1 or along the $\Delta t_{13}$ axis. The oscillation 
frequency is comparable to the inter-LL energy spacing and increases 
linearly with the magnetic field. The decay rate of these oscillations is 
comparable to the sum of the LL0 and LL1 dephasing rates extracted 
from Fig. \ref{LL1ExcitationDt13} (or Fig. \ref{LL1ExcitationTwoPulse}) 
for both positive and negative $\Delta t_{12}$. With increasing 
photoexcitation intensity, the  $\Delta t_{12}$ oscillations 
disappear quickly, even before the decay of the overall FWM signal 
changes significantly. In contrast, along $\Delta t_{13}$, 
 oscillations  start to appear with increasing intensity in both the 
doped and the undoped QWs, which  start to look alike as the 
photoexcited carriers dominate \cite{from-02-prb,dani}. 
After 
presenting the theory used to describe the ultrafast nonlinear 
optical response, 
we will discuss the physical origin of the 
 $\Delta t_{12}$ oscillations and show that they are mainly 
due to the X$\leftrightarrow$X+MP coherence.

\section{Theory Overview }

In this section we review  the theoretical formulation of the 
ultrafast nonlinear optical response of the 2DEG \cite{kara,chem-phys}. 
We start with the standard two--band 
Hamiltonian $H$ that describes conduction electron and valence hole 
discrete LLs coupled by the $e$--$e$, $e$--$h$, and $h$--$h$ Coulomb 
interactions \cite{schaf-book}. The coupling of the optical field 
is treated within the dipole approximation and induces interband transitions 
characterized by the Rabi energy $d(t)=\mu E(t) \sqrt{N}$, where $E(t)$ 
is the optical pulse, $\mu$ is the interband dipole transition matrix 
element, and  $N$ is the LL degeneracy
(we set $\hbar=1$ from now on). Similar to the experiment, 
we consider right--circularly polarized optical pulses ($\sigma_+$), 
which excite only spin--$\downarrow$ electrons due to the selection 
rules. 
For low 
photoexcitation intensity, the QH system is probed in a perturbative 
fashion, which allows for the observation of the 2DEG correlations. In this 
excitation regime, one can expand in powers of the optical 
field and describe the FWM signal by calculating the third--order 
nonlinear polarization. 

\subsection{Polarization and Population Equations of Motion} 

The ultrafast nonlinear optical response of the QH system is dominated 
by the interactions among its collective excitations: the interband 
excitons and the intraband magnetoplasmons. The LL$n$ electron--LL$m$ hole 
magnetoexcitons with total momentum ${\bf q}$ are created by the 
collective interband operators $\hat{X}^\dag_{{\bf q} nm}$ defined in  
Ref. \cite{chem-phys}. The  dipole optical transitions excite 
the LL$n$ magnetoexciton states $|X_n\rangle=\hat{X}^\dag_n|G\rangle$, 
where  $\hat{X}_n = \hat{X}^\dag_{nn0}$, whose difference from undoped
semiconductors is that the exciton operator now acts on the strongly 
correlated ground eigenstate, $|G\rangle$, of the many--body Hamiltonian 
$H$. For the small momenta relevant to optics experiments, the 2DEG 
intraband excitation spectrum is dominated by the collective MP modes. 
A LL$m\to$LL$n$ MP may be thought of as a pair formed by an electron in 
LL$n$ and a hole in the electron LL$m$ 2DEG created by the 
collective density operator 
$\hat{\rho}_{{\bf q}nm\sigma}^{e}$ \cite{QHE1,QHE2,chem-phys,oji,kallin-84}.  
It is convenient to also introduce a similar  operator
$ \hat{\rho}_{{\bf q}nm\sigma}^{h}$ for the valence band hole states 
\cite{chem-phys}. 
The photoexcited LL0 and LL1 optical transitions are dynamically coupled 
by the LL0 $\to$ LL1 inter--LL MPs.

Unlike in linear optics or Raman/inelastic light scattering experiments,  
the ultrafast FWM signal is 
generated by  interactions 
and correlations involving the elementary excitations 
of the system. The main challenge facing the calculation 
is the description of the interaction--induced 
quantum dynamics during time scales shorter than the dephasing and 
relaxation times. The interactions with  the ground state 
2DEG couple the exciton states to the X+2DEG$^*$ states 
$| Y_n\rangle = \hat{Y}^\dag_n | G \rangle$, where 2DEG$^*$ denotes an 
excited 2DEG configuration. These states are defined as follows after
requiring that $\langle X_m | Y_n \rangle=0$:
\begin{equation}
H |X_n \rangle = \Omega_n |X_n \rangle 
- ( 1 - \nu_n) \sum_{n' \neq n} V_{n'n}
|X_{n'} \rangle + |Y_n \rangle. \label{HonX}
\end{equation}
$\Omega_n$ is the $X_n$ energy, $V_{nn'}$ gives the static Coulomb--induced 
coupling of the different LL Xs, and $\nu_n$ is the ground state filling 
factor of the LL$n$ spin--$\downarrow$ electron states. The operator 
\begin{equation}
\label{Yop} {\hat Y}_n = [{\hat X}_n,H] - \Omega_n {\hat X}_n +
( 1 - \nu_n) \sum_{n' \neq n} V_{nn'} {\hat X}_{n'},
\end{equation}
then describes the interactions between  $X_n$ and all the 
other carriers in the system (i.e. X--X, X--MP, and X--2DEG interactions). 
By retaining for simplicity the contributions from the photoexcited LLs 
(LL0 and LL1), we can express ${\hat Y}_n$ as a linear combination of the 
operators $\hat{X}_{{\bf q} 01}\hat{\rho}^{e,h}_{{\bf q} nm}$ and 
$\hat{X}_{{\bf q} 10}\hat{\rho}^{e,h}_{{\bf q} nm}$ and obtain the property 
$\hat{Y} = \hat{Y}_1 = - \hat{Y}_0$ \cite{chem-phys}.  

The optical signal is determined by the LLn optical polarizations 
$P_n =\langle \hat{X}_{n}\rangle$, which satisfy the equations of motion 
\cite{chem-phys} 
\begin{equation}
i \partial_t P_n = \Omega_n  P_n 
- ( 1 - \nu_n) \sum_{n' \ne n} V_{nn'} P_{n'}
- d(t)[1 - \nu_n - \Delta \nu_{n} ]
+ \langle \hat{Y}_n \rangle, \label{eom-Pol} 
\end{equation} 
where the interband density matrix $\langle \hat{Y}_n \rangle$
gives the interaction--induced contribution that distinguishes 
semiconductors from atomic few level systems and 
determines the unusual off--resonant LL0 signal in the doped QW. 
 $\Delta\nu_n$
is the change in the LL$n$ filling factor induced by the 
photoexcitation and determines the Pauli Blocking (PSF) 
contribution to the nonlinear polarization $P_n$ 
(third term on the rhs of Eq.(\ref{eom-Pol})):  
\begin{equation} 
i \partial_t \Delta \nu_n = 
2 [d^*(t) P_n - d(t) P_n^*]/N
+ \langle [\hat{Y}_n, \hat{X}^\dag_n] \rangle^{*} 
- \langle [\hat{Y}_n, \hat{X}^\dag_n] \rangle. 
\label{eom-den} 
\end{equation} 
To describe  dephasing and  relaxation due 
to degrees of freedom not included in the Hamiltonian $H$, we also 
introduce the X dephasing rates $\Gamma_n$ and the population relaxation 
times $T_1$. 

The intraband density matrix 
$\langle [\hat{Y}_n, \hat{X}^\dag_n]  \rangle$
describes the redistribution of the photoexcited exciton 
populations due to 
the interactions and can be expressed as a linear combination of the 
density matrices $\langle \hat{X}^\dag \hat{X} \rangle$, 
$\langle \hat{\rho}^e_{\sigma} \hat{\rho}^e_{\downarrow} \rangle$, 
$\langle \hat{\rho}^e_{\sigma} \hat{\rho}^h_{\downarrow} \rangle$,
and $\langle\hat{\rho}^h_{\downarrow}\hat{\rho}^h_{\downarrow}\rangle$ 
\cite{chem-phys}. In the undoped system, one can show that the only 
independent intraband density matrices that contribute to the third 
order nonlinear polarization have the form 
$\langle \hat{X}^\dag \hat{X} \rangle$ \cite{chem-phys}. 
These describe exciton 
populations as well as X$\leftrightarrow$X coherences.
In the doped system,  the 
{\em ground state} electrons give an additional 
contribution to the third order polarization determined by the 
intraband density matrices
$\langle \hat{\rho}^e \hat{X}^\dag \hat{X} \rangle$, 
which in the undoped system contribute to fifth order ot higher. 
These density matrices describe the coherent coupling between the 
X and X+MP states, i.e. X$\leftrightarrow$X+MP many--particle 
coherences induced by the correlations of the QH system.

The main effects in the doped system 
can be described by calculating 
the interaction--induced density matrices 
$\langle [\hat{Y}_n, \hat{X}^\dag_n] \rangle$ and 
$\langle \hat{Y}_n \rangle$ up to third order in the optical fields. 
For this we first decompose these density matrices  
into correlated and uncorrelated parts, which allows 
us to isolate the contributions due to the 
correlations and devise appropriate approximations. Although in the 
undoped system this  can be achieved by introducing 
cumulants (the basis of the DCTS \cite{Axt}), in the doped 
system the ground state correlations as well as correlations between 
the photoexcited Xs and the 2DEG elementary excitations must also be 
taken into account. The DCTS assumptions and Wick's theorem break down 
in the case of the correlated 2DEG, so new schemes must be devised. 
To motivate a density matrix decomposition that reduces to the DCTS
in the case of undoped semiconductors but also applies to systems with 
strongly correlated ground states, we first 
decompose  the photoexcited 
many--body state $|\psi(t)\rangle$ that evolves from the ground state 
$| G \rangle$ into interacting and noninteracting 
parts as discussed in the next section.

\subsection{Decomposition of the photoexcited states}

We note that, since the Hamiltonian $H$ conserves the 
number of valence band holes, there is a one to one correspondence 
between the number of  holes created (annihilated) and the number of 
photons absorbed (emitted). We can therefore classify the photoexcited 
states in terms of the number of photoexcited holes:
$|\psi \rangle = |\psi_0 \rangle + |\psi_1 \rangle + |\psi_2\rangle
 + \cdots$,  
where $| \psi_n \rangle$ is the {\em collective} $n$--$h$ photoexcited 
state \cite{kara}. States with $n \ge 3$ do not contribute to the 
third--order polarization.

The linear response is described by the time evolution of the 1--$h$ 
many--body state $| \psi_{1L}\rangle$ calculated to first order in the 
optical field. We decompose this 1--$h$ state into X and 
interacting X+2DEG$^*$ 
 contributions:
\begin{equation}
\label{1hole} 
| \psi_{1L} \rangle = 
\sum_n \frac{P_{n}^{L}}{1 - \nu_n}| X_n \rangle +
| \bar{\psi}_{1L} \rangle,
\end{equation} 
where we require that $\langle X_n|\bar{\psi}_{1L}\rangle=0$. The X amplitude 
$P_n^{L} =  \langle X_n | \psi_{1L} \rangle$ coincides with the 
linear  polarization, while, to first order in the optical 
field, $\langle\hat{Y}_n \rangle=\langle Y_n|\bar{\psi}_{1L}\rangle$. 
The dephasing of the linear polarization is therefore determined by 
the time evolution of the X+2DEG$^*$ states $|Y_n\rangle$. These states 
can be expanded in a basis of states $| Y_{\alpha} \rangle$, such as 
the orthonormal Lanczos states introduced in Refs. \cite{kara,chem-phys} 
or the continuum of X+MP states 
$\hat{X}^\dag_{{\bf q} 01}\hat{\rho}^{e}_{{\bf -q} nm} |G\rangle$  
and $\hat{X}^\dag_{{\bf q} 10}\hat{\rho}^{e}_{-{\bf q} nm}|G \rangle$ 
for all values of the momentum ${\bf q}$ that contribute to $|Y \rangle$ 
\cite{chem-phys}. Retaining the above X+MP states corresponds to treating 
the  correlations between the X and a MP nonperturbatively, 
in a way analogous to 
the treatment of the carrier--magnon three--body correlations
in the case of the Hubbard 
\cite{hubb} or double exchange \cite{kapetan} Hamiltonians, 
which in 1D agreed well with exact results.
An analogous treatment of three--body correlations between 
Xs and Fermi sea excitations was used
to treat the Fermi Edge Singularity \cite{perakis-rev,fes}.  

To obtain the equation of motion for the linear X+MP amplitudes 
$\langle Y_{\alpha} | \bar{\psi}_{1L} \rangle$, we first consider the 
action of the Hamiltonian $H$ on $| Y_\alpha \rangle$ and define the 
state $| Z_{\alpha} \rangle$ by requiring that it is orthogonal to both 
$| Y_\alpha \rangle$ and  all  $| X_n \rangle$ \cite{kara,chem-phys}:
\begin{equation} 
H | Y_\alpha \rangle = \bar{\Omega}_{\alpha} 
| Y_\alpha \rangle + \sum_{n} W_{\alpha n} | X_n \rangle 
+ | Z_{\alpha} \rangle, 
\end{equation} 
where $\bar{\Omega}_{\alpha}$ is the energy of the state
$|Y_{\alpha}\rangle$, $W_{\alpha n}$ gives the interaction--induced 
coupling between $ |Y_{\alpha} \rangle$ and  $|X_{n} \rangle$, and 
$| Z_{\alpha} \rangle$ comes from the scattering between the X and 
the MP. We thus obtain the equation of motion \cite{kara,chem-phys} 
\begin{equation} 
i \partial_{t} 
\langle Y_{\alpha} | \bar{\psi}_{1L} \rangle
=  (\bar{\Omega}_{\alpha} - i \gamma_{\alpha}) 
\langle Y_{\alpha} | \bar{\psi}_{1L} \rangle
+ \sum_{n} W_{\alpha n}^* P_{n}^L
+ \langle Z_{\alpha} |   \bar{\psi}_{1L} \rangle, 
\end{equation} 
whose solution has the form 
$\langle Y_{\alpha} | \bar{\psi}_{1L} \rangle = \sum_{n} W_{\alpha n}^* 
\int_{- \infty}^{t} K_{\alpha}(t - t') P_{n}^L(t')dt'$.
The X+MP correlation function 
$K_{\alpha}(t) = -i\langle Y_{\alpha}|e^{- i \bar{H}t}|Y\rangle$  
describes memory effects governed by noninstantaneous 
X+MP interactions, where 
 $\bar{H}$ 
is the Hamiltonian $H$ projected within the subspace of X+2DEG$^*$ 
states.
The semiclassical 
approximation  corresponds to approximating 
$K_{\alpha}(t)  \sim i T_{2}^{-1}  \delta(t)$, where $T_2$ 
is the exciton dephasing time. However, 
 memory effects described by 
the time dependence of 
$K(t)$
are important 
for determining the LL1 exciton linewidth for $\nu <2$ (Fig. \ref{LinAbsDoped})
\cite{from-02-prb,kara,chem-phys}.

The 2--$h$ and 0--$h$ many--body states $| \psi_2 \rangle$ and 
$| \psi_0 \rangle$ are photoexcited via two--photon nonlinear processes. 
The first photon excites a 1--$h$ state, $|X_n\rangle$ or 
$|\bar{\psi}_{1L}\rangle$, from the ground state. The second photon 
then excites ($|\psi_2\rangle$) or deexcites ($|\psi_0\rangle$) a second 
$e$--$h$ pair. This second transition may or may not be accompanied by 
interactions with the carriers in the 1--$h$ states already excited by 
the first transition. We separate out the above interacting and 
noninteracting contributions to second order in the optical field as follows: 
\begin{equation}
 | \psi_2\rangle =  \frac{1}{2} \sum_{nm} \frac{P_{n}^L
P_{m}^L}{(1 - \nu_n )(1 - \nu_{m})}  
 \hat{X}^\dag_{n} | X_{m} \rangle + 
 \sum_{n} \frac{P_{n}^L}{1 - \nu_n} {\hat X}^{\dag}_{n} |
\bar{\psi}_{1L} \rangle + |\bar{ \psi}_2\rangle,
\label{2hole}
\end{equation}
where $|\bar{ \psi}_2\rangle$ describes the correlated X-X and 
X-X+2DEG$^*$ contributions, and 
\begin{equation}
| \psi_0 \rangle = \langle G | \psi \rangle \, | G\rangle - 
\sum_{n} \frac{P_{n}^{L*}}{1 - \nu_n}
{\hat X}_{n} | \bar{\psi}_{1L} \rangle + |
\bar{\psi}_0 \rangle, \label{0hole} 
\end{equation}
where $|\bar{\psi}_0 \rangle$ is the  2DEG$^*$ state created by the 
Raman process of excitation and then de-excitation 
of an $e$--$h$ pair assisted by correlations with the 2DEG. 

\subsection{Density matrix decompositions} 

By substituting the  above decomposition of the photoexcited 
states,
one arrives at the following decomposition of the density matrix
$\langle \hat{M} \rangle$, where the operator $\hat{M}$, 
$\langle G | \hat{M} | G \rangle = 0$, connects states with the same 
number of holes:
\begin{eqnarray}  
\label{Mc-states} 
\langle \hat{M} \rangle 
&=& 
\langle \hat{M} \rangle_c
+ \sum_n \frac{P_n^{L*}}{1 - \nu_n}
\langle G | [ \hat{X}_n, \hat{M}] | \bar{\psi}_{1L} \rangle 
+ \sum_n \frac{P_n^{L}}{1 - \nu_n}
\langle \bar{\psi}_{1L} | [ \hat{M}, \hat{X}_n^\dag] | G \rangle 
\nonumber \\
&&
+ \sum_{nm} \frac{P_n^{L*} P_m^L}{(1 - \nu_n) (1 - \nu_m)} 
\langle X_n | \hat{M} | X_m \rangle 
+ O(E^4), 
\end{eqnarray} 
where the correlated contribution $\langle \hat{M} \rangle_c$ is 
determined by its equation of motion as discussed in Ref. \cite{chem-phys}.
The rest of the terms in Eq.(\ref{Mc-states})
are given by products of 1--$h$ state amplitudes, 
X or X+2DEG$^*$, and arise from uncorrelated interband transitions.
The time evolution of these product terms is determined by the X and 
X+2DEG$^*$ dephasing. In contrast, the dephasing of the correlated 
contribution $\langle \hat{M} \rangle_c$ is determined by the intraband 
dynamics, which introduces new time scales that affect the evolution 
of the transient 
FWM signal.

In the undoped system, the ultrafast 
dynamics is solely determined  by the X states,  
whose coherent coupling  is described 
by the equation of motion
 \cite{chem-phys}
\begin{eqnarray} 
&&
i \partial_t
\langle | X_m \rangle \langle X_n | \rangle_c= 
\left(\Omega_n - \Omega_m - i \gamma_{nm} \right) 
\langle | X_m \rangle \langle X_n | \rangle_c
\nonumber \\ 
&& \qquad
+ (1-\nu_m) \sum_{m' \ne m} V_{m'm}
\langle | X_{m'} \rangle \langle X_n | \rangle_c
- (1-\nu_n) \sum_{n' \ne n} V_{nn'} 
\langle | X_m \rangle \langle X_{n'} | \rangle_c
\nonumber \\
&& \qquad
+ i P_n^L P_m^{L*} (\Gamma_n + \Gamma_m - \gamma_{nm}) 
+ \langle | X_m \rangle \langle Y_n | \rangle_c
- \langle | X_n \rangle \langle Y_m | \rangle_c^*,
\end{eqnarray}  
 As can be seen from the above 
equation, incoherent exciton populations ($n=m$) or X$\leftrightarrow$X 
coherences ($n\ne m$) can be photoexcited due to (i) the difference 
between the intraband relaxation rate $\gamma_{nm}$ and the sum of the 
exciton dephasing rates $\Gamma_n + \Gamma_m$, and  (ii) the coupling 
between the X and  X+MP states described by the last two terms on the rhs.
The former is the only source in the undoped system, while the 
latter X$\leftrightarrow$X+MP coupling  dominates in the doped system and 
is described by the equation of motion \cite{chem-phys}
\begin{eqnarray}
&& 
i \partial_t\langle |X_m \rangle \langle Y_{\alpha} | \rangle_{c} =
\left(\bar{\Omega}_{\alpha} - \Omega_m - i \Gamma_{\alpha m} \right) 
\langle |X_m \rangle \langle Y_{\alpha} | \rangle_{c} 
+ \sum_{n'} W_{\alpha n'}^{ *} 
\langle | X_{m} \rangle \langle X_{n'} | \rangle_{c} 
\nonumber \\
&&  \qquad
+ ( 1 - \nu_m) \sum_{m' \ne m} V_{m'm} 
\langle |X_{m'} \rangle \langle Y_{\alpha} | \rangle_{c} 
+ i \left( \Gamma_m + \gamma_{\alpha} -  \Gamma_{\alpha m} \right) 
P_m^{L*}  \langle Y_{\alpha} | \bar{\psi}_{1L} \rangle 
\nonumber \\
&& \qquad
+ \langle |X_m \rangle \langle Z_{\alpha} | \rangle_{c} 
- \left[  \langle |Y_{m}  \rangle \langle Y_{\alpha} | \rangle_{c}
- \langle \bar{\psi}_{1L} | Y_{m} \rangle  
 \langle Y_{\alpha} | \bar{\psi}_{1L} \rangle \right].  
\end{eqnarray}  
The separation of the photoexcited states into 
interacting and noninteracting arts 
also suggests the following decomposition of the 
interband density matrix:
\begin{eqnarray}
\label{interactions} 
\langle {\hat Y} \rangle 
&=& 
\sum_n \frac{P_{n}^{L*}}{1-\nu_n}
\langle G|[\hat{X}_n,\hat{Y}] |\psi_2\rangle 
+ \sum_n \frac{P_{n}^{L}}{1 - \nu_{n}}  
\langle [\hat{Y},\hat{X}^\dag_{n}] \rangle_c  
\nonumber \\ 
&&
+ \frac{1}{2} \sum_{nm} 
\frac{P_{n}^L \, P_{m}^L}{( 1 - \nu_{n}) ( 1 - \nu_{m})}   
\langle \bar{\psi}_{1L} | 
[[\hat{Y},\hat{X}_n^\dag],{\hat X}^{\dag}_{m}] 
| G \rangle + \langle \hat{Y} \rangle_c, 
\end{eqnarray}
where the correlated part $\langle \hat{Y} \rangle_c $ is 
determined 
by the dynamics of the states $|Y\rangle$ and $|\bar{\psi}_0 \rangle$,
and by the coupling between $|\bar{\psi}_1\rangle$ and $|\bar{\psi}_2\rangle$
(exciton--biexciton transition amplitude in undoped semiconductors 
\cite{Axt}). The first term on the rhs of Eq.(\ref{interactions}) 
describes the coherent X--X interaction contribution,
determined by the dynamics of the  
2--$h$ state $[ \hat{Y}^\dag,  \hat{X}_n^\dag ] | G \rangle$ which 
is a linear combination of  two--exciton states 
\cite{kara,chem-phys}. In the 
undoped system, the above 2--$h$ amplitude can be expressed in terms of the 
X--X correlation function discussed in Ref. \cite{ost-98}, 
which is analogous to 
the X--MP correlation function $K_{\alpha}(t)$ 
discussed above.  The 
rest of the terms on the rhs of Eq.(\ref{interactions}) are due to the 
scattering of the polarization $P_n^L$ off incoherent populations and 
intraband coherences, such as the X$\leftrightarrow$X and 
X$\leftrightarrow$X+MP coherences discussed above. Noting the huge 
difference in the LL0 signal between the doped and undoped QWs, we 
conclude that the contribution due to the X$\leftrightarrow$X+MP coherences 
dominates the ultrafast nonlinear optical response of the 2DEG
in the femstosecond regime.  In the next 
section we present a model calculation based on the above microscopic 
theory that captures the main experimental features. 

\section{ Comparison between theory and experiment} 

In this section we discuss the results of our numerical calculation 
of the transient three--pulse  FWM spectrum at $\nu=1$. 
These were 
 obtained as 
above  after retaining the states 
$| X_{n} \rangle$, $n=0,1$, and $|Y \rangle $ and treating 
for simplicity the effects 
of the rest of the basis states by introducing dephasing rates 
\cite{kara,chem-phys}. The independent parameters that enter this model 
calculation were estimated by comparing to the experimental linear 
absorption \cite{from-02-prb,kara,chem-phys}; our conclusions are not 
sensitive to their precise values. Here we consider the ideal 2D system, 
where the coherent MP contribution discussed in Refs. \cite{kara,chem-phys} 
vanishes. 

\begin{figure}[t]
\begin{center}
\includegraphics[width=1.0\textwidth]{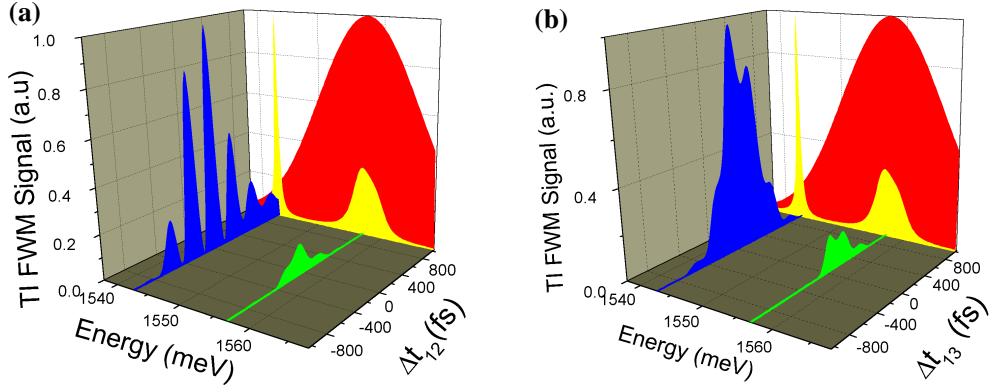}
\end{center}
\caption[]{Calculation of the FWM Signal in the doped QW along
(a) the $\Delta t_{12}$ axis
(b) the $\Delta t_{13}$ axis.
Back panel: linear absorption spectrum and optical pulse. 
}
\label{theory}
\end{figure}

\begin{figure}[t]
\begin{center}
\includegraphics[width=0.8\textwidth]{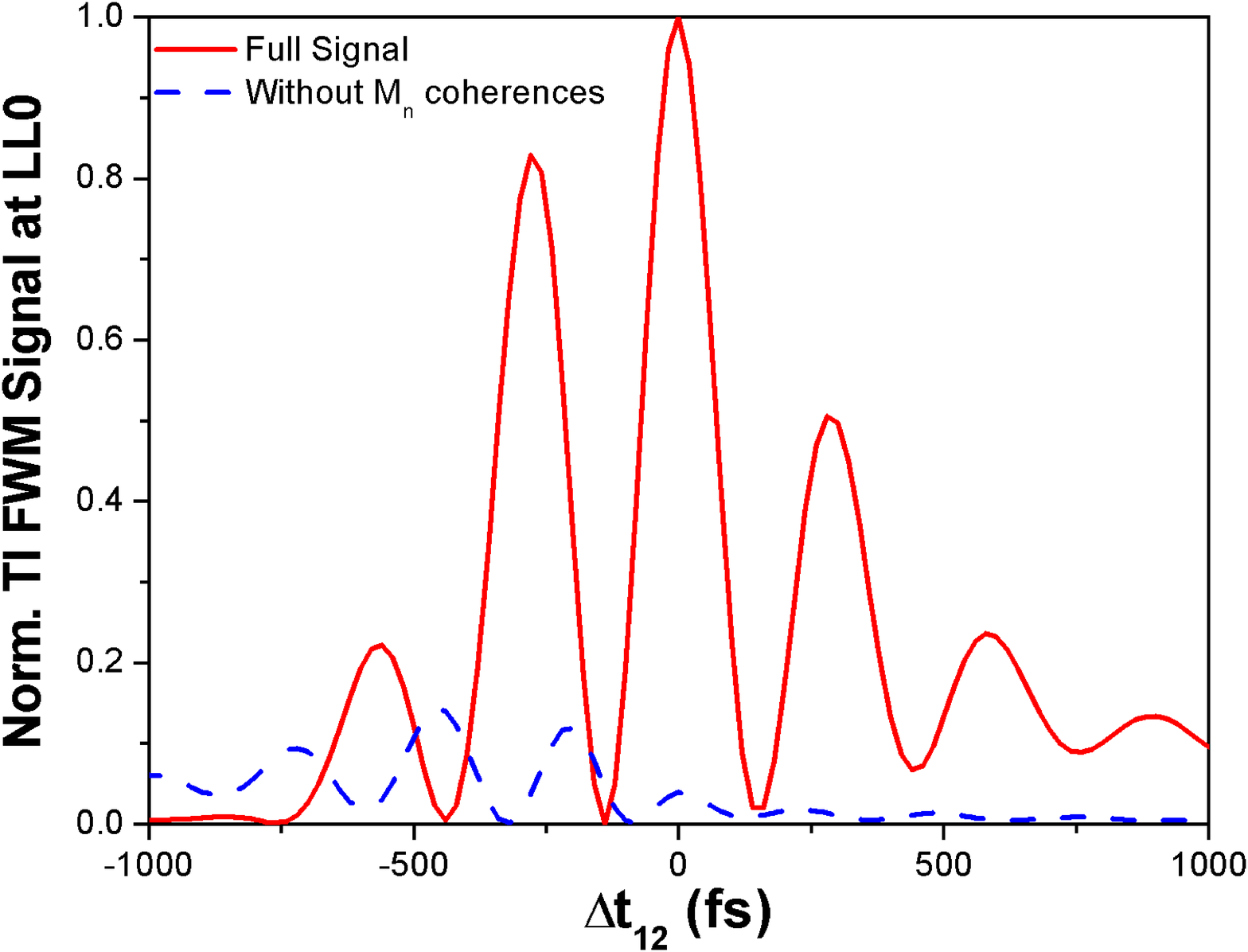}
\caption{\label{terms}
Comparison of the calculated LL0 FWM signal along the $\Delta t_{12}$ 
axis (solid line) with the signal calculated by setting $M_n=0$ 
(dashed line).
}
\end{center}
\end{figure}

Fig. \ref{theory} shows the calculated transient FWM
for LL1 photoexcitation, 
 while the back
panel shows the calculated linear 
absorption. These results  reproduce the qualitative temporal and spectral 
features observed in the experiment within the
femstosecond coherent 
regime, as function of energy, the two time delays, and the central 
photoexcitation frequency. The main contribution to the calculated FWM 
signal comes from the X$_{n} \rightarrow X_{01}$+MP  
coherences $M_n= \langle |X_n\rangle \langle Y|\rangle_c$. This can 
be seen in Fig. \ref{terms}, which compares to the 
calculation with $M_n=0$. In the latter case, the FWM signal is determined 
by PSF and X--X contributions analogous to the undoped system, 
whic, similar to 
the experiment,   
give a small LL0  signal. 
 The 
X$\leftrightarrow$X+MP coherences can be photoexcited either by pulses 
1 and 3 ($M_{n}^{13}$) or by pulses 2 and 3 ($M_n^{23}$). 
$M_0$ ($M_1$) comes from the scattering of the LL0 valence 
hole (LL1 conduction electron) to LL1 (LL0) accompanied by the emission 
of an inter--LL MP.

\begin{figure}[t]
\begin{center}
\includegraphics[width=1.0\textwidth]{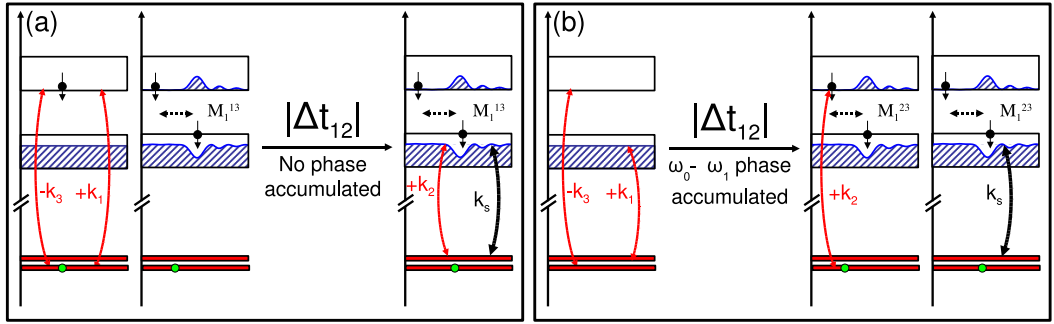}
\caption{\label{schematic}
Third-order process that describes the main contribution to the 
FWM signal due to
(a) $M_1^{13}$
(b) $M_1^{23}$.
 } 
\end{center}
\end{figure}

The main contribution to the 
$\Delta t_{12}$ oscillations of the FWM signal comes from the 
interference between the two  
nonlinear processes described  schematically in Fig. \ref{schematic}. 
Pulses ${\bf k}_1$ and ${\bf k}_3$ arrive simultaneously in the sample 
to create a density of excitons in LL1. These excitons scatter into LL0 
with the excitation of a MP, thereby creating the coherence $M_1^{13}$. 
This coherence then 
evolves for a time $|\Delta t_{12}|$ accumulating negligible 
phase due to the small $M_1$ energy. It is then probed by a $P_0$ 
polarization created by ${\bf k}_2$, resulting in a FWM signal in 
${\bf k}_1+{\bf k}_2-{\bf k}_3$ (Fig. \ref{schematic}a). Due to the symmetry 
of ${\bf k}_1$ and ${\bf k}_2$ in the ${\bf k}_1+{\bf k}_2-{\bf k}_3$ signal, 
we also have a process where ${\bf k}_2$ and ${\bf k}_3$ create the 
coherence $M_1$ (i.e. $M_1^{23}$), which is then probed by ${\bf k}_1$. 
However, one must keep track of the time delays. As shown in Fig. 
\ref{schematic}b, ${\bf k}_1$ and ${\bf k}_3$ arrive together
and contribute a LL1 polarization ( ${\bf k}_3$) 
and  a LL0 polarization (${\bf k}_1$). 
These polarizations evolve in the sample for a time $|\Delta t_{12}|$ and 
accumulate a phase of $(\Omega_0-\Omega_1)\Delta t_{12}$, where $\Omega_n$ 
denotes the energy of $X_n$. Pulse ${\bf k}_2$ then contributes a LL1 
polarization that creates the $M_1^{23}$ coherence with the decaying LL1 
polarization from ${\bf k}_3$. $M_1^{23}$ is instantaneously probed by the 
decaying LL0 polarization created earlier by ${\bf k}_1$, which  results 
in a FWM signal with the accumulated phase $(\Omega_0-\Omega_1)\Delta t_{12}$. 
The above contributions from 
$M_1^{13}$ and $M_1^{23}$ will interfere with each other
for $|\Delta t_{12}|$ within the decay times of the polarizations, 
which results in the 
strong oscillations at the inter-LL frequency $\Omega_1-\Omega_0$ along the 
$\Delta t_{12}$ axis observed in the experiment. 
At the same time, 
the symmetric temporal profile along the 
$\Delta t_{13}$ axis of the  LL0 FWM signal due to $M_n$  
results from the much larger LL1 polarization dephasing as compared to 
the LL0 polarization dephasing,  due to the 
 $X_1 \rightarrow X_{01} + MP$ scattering process.

\section{Conclusions} 

In conclusion,  we discussed recent ultrafast two-- and 
three--pulse FWM results that demonstrate the important 
role of  {\em non--instantaneous} correlations between 
 photoexcited excitons and the inter--LL collective excitations of 
the 2DEG. We showed that three--pulse transient FWM 
spectroscopy can be used to access simultaneously the intra-- and 
inter--band coherent dynamics of the QH system. 
Even for very small excitation of the LL0 transition, the FWM signal 
in the doped system is dominated by a large off--resonant peak at the LL0 
energy with strong coherent oscillations and symmetric 
temporal profile.
 Using a microscopic many-body theory we 
showed that this signal is due to many-particle coherences created via the 
non-instantaneous interactions of photoexcited carriers and MPs. 
In particular, the noninstantaneous 
 X$_1 \rightarrow$X$_{01}$+MP interaction 
process both creates  an intraband coherence
and leads to strong LL1 exciton dephasing. Such effects 
govern the LL0 FWM temporal and spectral profiles.
We showed for example that strong temporal oscillations 
result from the interference of different FWM contributions 
of the above  intraband coherences. 
The combination of ultrafast nonlinear spectroscopy 
and QH physics initiates a new field of  QH dynamics. Future 
experimental and theoretical activity in this  area will further progress 
our understanding and manipulation of non--equilibrium correlations and 
quantum coherent phenomena 
  in nanostructures.

This work was supported by the EU Research Training Network  HYTEC 
(HPRN-CT-2002-00315) and by the U.S. Department of Energy under 
Contract No. DE-AC03-76SF00098. Part of this work was performed in 
collaboration with N. Fromer and A. T. Karathanos.

\end{document}